\documentclass{jfm}
\usepackage{graphicx}
\usepackage{epstopdf, epsfig}
\usepackage{times,mathptmx,amsmath,amssymb}

\shorttitle{Dipolophoresis in concentrated suspensions}
\shortauthor{S. Mirfendereski and J. S. Park}

\title{Dipolophoresis in concentrated suspensions of ideally polarizable spheres}

\author{Siamak Mirfendereski\aff{1}
   and
  Jae Sung Park\aff{1}
  \corresp{\email{jaesung.park@unl.edu}}
  }

\affiliation{\aff{1}Department of Mechanical and Materials Engineering, University of Nebraska-Lincoln,
Lincoln, NE 68588, USA}

\begin{document}

\maketitle

\begin{abstract}

The dynamics of ideally polarizable spherical particles in concentrated suspensions under the effects of nonlinear electrokinetic phenomena is analysed using large-scale numerical simulations. Particles are assumed to carry no net charge and considered to undergo the combination of dielectrophoresis and induced-charge electrophoresis termed dipolophoresis. Chaotic motion and resulting hydrodynamic diffusion are known to be driven by the induced-charge electrophoresis, which dominates the dielectrophoresis. Up to a volume fraction $\phi \approx 35\%$, the particle dynamics seems to be hindered by the increase in the magnitude of excluded volume interactions with concentration. However, a non-trivial suspension behaviour is observed in concentrated regimes, where the hydrodynamic diffusivity starts to increase with a volume fraction at $\phi \approx 35\%$ before reaching a local maximum and then drastically decreases as approaching random close packing. Similar non-trivial behaviours are observed in the particle velocity and number-density fluctuations around volume fractions that the non-trivial behaviour of the hydrodynamic diffusion is observed. We explain these non-trivial behaviours as a consequence of particle contacts, which are related to the dominant mechanism of particle pairings. The particle contacts are classified into attractive and repulsive classes by the nature of contacts, and in particular, the strong repulsive contact becomes predominant at $\phi > 20\%$. Moreover, this transition is visible in the pair distribution functions, which also reveal the change in the suspension microstructure in concentrated regimes. It appears that strong and massive repulsive contacts along the direction perpendicular to an electric field promote the non-trivial suspension behaviours observed in concentrated regimes.
\end{abstract}

\section{Introduction}\label{sec:introduction}
The collective dynamics of concentrated suspensions undergoing electrokinetic and hydrodynamic interactions remains relatively unexplored as compared to that of dilute and semi-dilute suspensions. Beyond the semi-dilute limit, the use of the electrokinetics is limited by the poor understanding of its effect on suspension dynamics in semi-dilute and concentrated regimes, where multi-body interactions become more significant \citep{Russel1991, Bazant2009}. As one example, suspensions of dielectric particles undergoing dielectrophoresis (DEP) or electrorheological (ER) fluids consist of solid particles dispersed in a non-conducting fluid, exhibiting a dramatic viscosity enhancement that is reversible and controlled by applied electric fields \citep{Pohl1978, Von2002, Sheng2012}. It is due to the rapid formation of particle chains/columns along the applied field direction as a result of dipolar interactions between particles. The rheological and yield stress responses of such suspensions or ER fluids have been well documented with the volume fractions in the range of 0.05 - 0.30 \citep{Liu2012}. However, these responses have yet to be fully studied with volume fractions beyond this range. It is worth noting that there is a recent study that uses a thermodynamic theory for directed assembly of dielectric particles at the concentrated regime \citep{Sherman2018}.

If particles are polarizable and placed in a conducting fluid, the dynamics of such suspension becomes even complex as they can acquire an additional non-uniform surface charge. The additional surface charge results in a non-linear fluid flow around the particles, leading to the induced-charge electrophoresis (ICEP) \citep{squires2004, Squires2006}. The experimental and numerical studies to date have focused on suspensions undergoing ICEP at solid volume fractions $\phi$ of only up to 0.15 for spherical particles \citep{Park2010} or at effective volume fractions $nl^3$ of up to 1 for rigid fibres (where $n$ is the number density and $l$ the fibre half-length) \citep{Saintillan2006, Rose2009}. At dilute and semi-dilute regimes, the suspension undergoing ICEP displays a transient pairing dynamics that particles attract along the direction of the electric field, pair up, and separate in the transverse direction. However, the high complexity of the suspension dynamics undergoing ICEP is anticipated at semi-concentrated and concentrated regimes, possibly yielding unexpected or undesirable effects. Yet, no such suspensions, to our knowledge, have been explored until now.

It is important to note that when polarizable particles are placed in a conducting fluid, particle motions undergoing both ICEP and DEP occur concurrently, which is sometimes referred to as dipolophoresis (DIP) \citep{Shilov1981}. For ICEP, the particle velocity due to induced-charge fluid flow is easily seen by the Helmholtz-Smoluchowski equation to scale quadratically with the magnitude of the applied electric field $E = |\textbf{\textit{E}}|$ because the additional surface charge or induced zeta potential is driven by the electric field. A self-consistent calculation of the resulting particle motions also requires accounting for the Maxwell electric stress in the fluid, which is the same order $O(E^2)$ as for the hydrodynamic stresses generated by ICEP. The Maxwell electric stress accounts for DEP forces and torques. This concurrence can also be explained by symmetry breaking. Even in a uniform electric field, the presence of several polarizable particles such as a suspension disturbs a local electric field around the particles and then generates a non-uniform electric field. This symmetry breaking of the electric field in turn results in relative motions between particles due to DEP and ICEP. Note that both DEP and ICEP effects on particle motions indeed disappear in the single sphere placed in a uniform electric field due to fore-aft symmetry. In general, DIP interactions are governed by ICEP in a suspension because the leading-order contributions to ICEP and DEP interactions are a Stokes dipole and a potential quadrupole, respectively, where the former has a slower decay with separation distance $R$ as $O(R^{-2})$ than the later as $O(R^{-4})$ \citep{Saintillan2008, Park2010}. That is, the particles undergo random chaotic motions that constantly cause particle configurations to rearrange, but do not lead to the formation of chains as in the case of DEP only. 

In this paper, we use large-scale numerical simulations to investigate the suspension dynamics of non-colloidal, polarizable spheres in a uniform electric field for concentrations up to random close packing. An effective simulation method for excluded volume interactions facilitates the study of highly concentrated suspensions. The simulation details are reported in \S \ref{sec:Simulation_method}. The simulation results are presented \S \ref{sec:results}, where the analysis of pair interactions is used to define different natures of particle contacts. We conclude in \S \ref{sec:conclusion}.


\section{Simulation method}\label{sec:Simulation_method}
We consider a suspension of $N$ identical neutrally buoyant ideally polarizable spheres of radius $a$ in a viscous electrolyte with the viscosity $\eta$ and permittivity $\varepsilon$. A cubic periodic domain is used to simulate an unbounded infinite suspension. An external uniform electric field $\textbf{\textit{E}}_0 = E_0\hat{\mathbf{z}}$ is applied in the $z$-direction. The particles are assumed to carry no net charge so the linear electrophoresis is not expected to occur. We also assume thin Debye layers, weak electric fields, and zero Dukhin number for no surface conduction  \citep{squires2004}.

For electrokinetic and hydrodynamic interactions in a suspension, we adopt the simulation method developed in our previous work \citep{Park2010}, for which a new approach was introduced to efficiently prevent particle overlaps. The outline of the method is as follows. Based on pair interactions due to DIP \citep{Saintillan2008}, the translational velocity $\dot{\textbf{\textit{x}}}_\alpha$ of a given particle $\alpha$ in a suspension can be expressed by:
\begin{equation}
{\dot{\textbf{\textit{x}}}}_{\alpha} = \frac{\varepsilon a E_{0}^{2}}{\eta} \sum_{\beta=1}^{N} \left[\mathsfbi{M}^{DEP}(\textbf{\textit{R}}_{\alpha \beta}/a)+\mathsfbi{M}^{ICEP}(\textbf{\textit{R}}_{\alpha \beta}/a)\right] \boldsymbol{:}\hat{\textbf{\textit{z}}}\hat{\textbf{\textit{z}}}, \ \ \ \ \ \alpha=1,...,N,\label{eq:sums}
 \end{equation}
where $\textbf{\textit{R}}_{\alpha \beta} = \textbf{\textit{x}}_{\beta} - \textbf{\textit{x}}_{\alpha}$ is the separation vector between the particle $\alpha$ and particle $\beta$, and $\mathsfbi{M}^{ICEP}$ and $\mathsfbi{M}^{DEP}$ are third-order dimensionless tensors accounting for the DEP and ICEP interactions, respectively. It is shown that these two tensors are entirely determined by these scalar functions of the dimensionless inverse separation distance $\lambda= 2a/\vert \textbf{\textit{R}} \vert$. For far-field interactions ($\lambda \ll 1$), the DEP and ICEP far-field interaction tensors can be computed using the method of reflections and expressed to order $ O(\lambda^4)$ in terms of two fundamental solutions of Stokes equations $\mathsfbi{S}$ and $\mathsfbi{T}=\nabla^{2}\mathsfbi{S}$, which are the Green's functions for a Stokes dipole and for a potential quadrupole, respectively \citep{Kim2013}. These tensors can be expressed as the periodic version of the far-field tensors to account for far-field interactions between particles $\alpha$ and $\beta$ and all its periodic images, which are valid to order $O(R_{\alpha\beta}^{-4})$. Since a high-order computation $O(N^2)$ is required to direct calculation of these sums in equation~(\ref{eq:sums}), the smooth particle mesh Ewald (SPME) algorithm based on the Ewald summation formula of \citet{hasimoto1959} and on fast Fourier transforms can be used to accelerate the calculation of the sums to $O(N \log N)$ operations \citep{Park2010}. However, the near-field corrections are necessary when particles are close to each other (typically $|\textbf{\textit{R}}_{\alpha\beta}| < 4a$) as the method of reflections becomes inaccurate. This is achieved by correcting the far-field tensor with a more accurate method of twin multiple expansions \citep{Saintillan2008}. The method of twin multiple expansions is very accurate down to separation distances on the order of $|\textbf{\textit{R}}_{\alpha\beta}|\approx 2.005 a$ \citep{Park2010}.

To prevent particle overlaps that occur due to the use of finite time steps in simulations, the application of a repulsive interparticle force is necessary. The contact algorithm previously developed by \citet{Park2010}, which was successful up to a semi-dilute regime without introducing any unphysical long-range interactions, is found to fail beyond a volume fraction of 0.20 for this study. To this end, an effective algorithm functionally identical to the potential-free algorithm \citep{melrose1993simulations} is used to prevent excessive particle overlaps due to DIP interactions, where particles are moved almost exactly, within round off errors $(\sim 2.005a)$, to contact. The form of the repulsive potential for excluded volume ($EV$) interactions is:
\begin{equation}
   U^{EV}=\frac{1}{2}k\left ( 2a-R \right )^2,
\end{equation}
where $k$ is the time-step-dependent prefactor, which can be expressed as $k = 3\pi \eta a /\Delta t$ \citep{Sherman2018}. This potential contributes to the displacement along the direction connecting the center of the two spheres at the points of closest approach. The robustness with respect to the excluded volume interactions were tested by using different values of the prefactor in equation (2.2) and no excluded volume interactions, where almost identical results were produced. Another troublesome for suspension simulations at concentrated regimes is the initial configuration of random particle distributions. Here, the initial random configurations were generated using a similar procedure to ones suggested for dense hard-sphere systems \citep{rintoul1996computer,clarke1987numerical,stillinger1964systematic,jodrey1985computer}. All runs were started with hard-sphere equilibrium configurations, but the first 100 time runs were discarded for better steady-state configurations beginning to compute averages.

In the remainder of the paper, all variables are made dimensionless using the following characteristic length and velocity scales: $l_{c}=a$ and $u_{c}=\varepsilon a E_{0}^{2} / \eta$.

\section{Results and discussion\label{sec:results}}
\begin{figure}
  \centerline{
  \includegraphics[width=5.2 in]{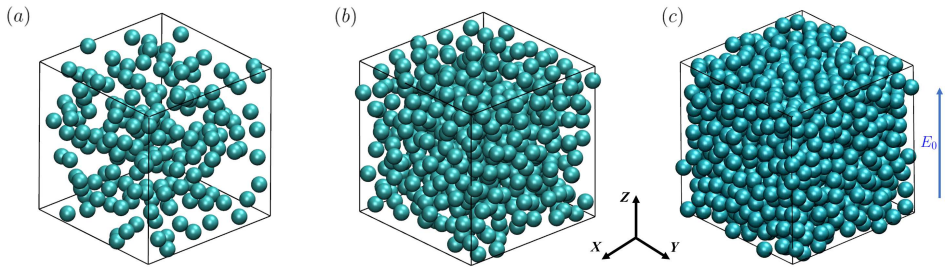}}
  \caption{Snapshots of particle distributions in a periodic cell of dimensions $L_{x}\times L_{y}\times L_{z}=20^{3}$ at different volume fractions: $(a)$ $\phi=10\%$ (dilute), $(b)$ $\phi=25\%$ (semi-dilute), and $(c)$ $\phi=59\%$ (concentrated). Also see the accompanying online movies.}\label{fig1}
\end{figure}

Large-scale simulations are performed in a periodic domain using the algorithm described above. The particle distributions in a suspension undergoing DEP and ICEP at different volume fractions are shown in figure \ref{fig1} and can also be seen in accompanying online movies. Particles are found to exhibit a random chaotic motion due to a dominant effect of ICEP over DEP as reported in \citet{Park2010}. The particle dynamics is governed by two mechanisms -- the particle-particle interactions originated from DIP and the excluded volume (steric) interactions. As clearly seen in the accompanying online movies, the combination of these two mechanisms leads to transient clusters, which constantly break up and form over time. \citet{Park2010} showed that in a dilute regime (weak steric interactions), particles tend to be attracted along the field direction, briefly pair up, and separate near the transverse direction. The particle dynamics is severely hindered for concentrated suspensions owing to strong steric interactions with quite a few neighbouring particles (figure \ref{fig1}$c$).

\subsection{Hydrodynamic dispersion and velocity statistics}
\begin{figure}
  \centerline{
  \includegraphics[width=4.9in]{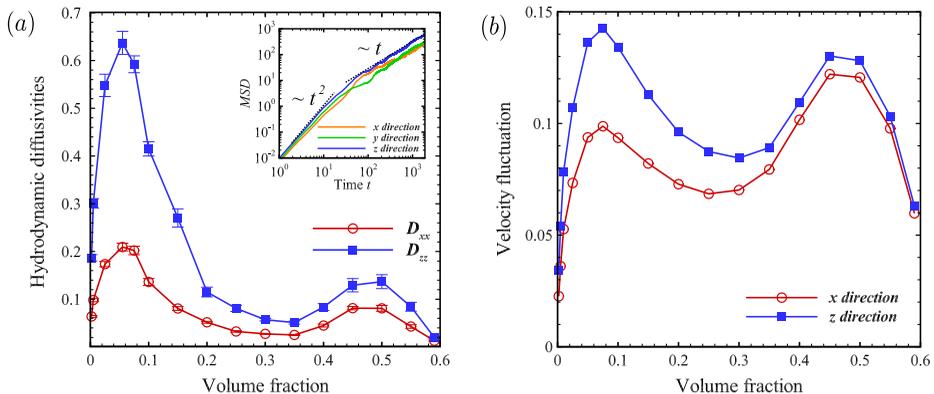}}
  \caption{$(a)$ Hydrodynamic diffusivities in the transverse direction ($x$) and the field direction ($z$) versus volume fractions with the error bars based on the block-averaging method \citep{Flyvbjerg1989}. Inset: mean-square displacement (MSD) curves in the transverse ($x$, $y$) and field direction ($z$) at $\phi = 50\%$ as functions of time on a log-log scale. $(b)$ Standard deviations of particle velocities as functions of volume fraction in the $x$ and $z$ directions.}\label{fig2}
\end{figure}

For hydrodynamic dispersion, the mean-square displacements (MSD) versus time are calculated to quantify the hydrodynamic diffusion of a suspension undergoing DIP. Initially, the MSD curve exhibits a quadratic growth with time. After a few particle-particle interactions, a transition to the diffusive regime takes place for the particle motion with linear growth of the mean-square displacement. In the inset of figure \ref{fig2}$(a)$, the log-log plot of mean-square displacements at $\phi = 50\%$ is presented to clearly show the diffusive behavior in concentrated regimes, where a transition from a ballistic regime ($\sim t^2$) to a diffusive regime ($\sim t$) is observed. The average slopes of the MSD curves at the diffusive regime indicates an effective hydrodynamic diffusivity tensor $\mathsfbi{D}$. It was previously reported that the hydrodynamic diffusion is primarily a consequence of ICEP interactions causing the rearrangement of particle configuration with no long-lasting clusters, while DEP interactions result in the formation of stable chains and trapping the particles in larger structures, leading to a sub-diffusion \citep{park2011,Park2011sm}. 

Figure \ref{fig2}$(a)$ shows the dependence of the diffusivity on the volume fraction from dilute to concentrated regimes. For the error bars on the plot, the block-averaging method is used to calculate the standard errors of the time-averaged quantity \citep{Flyvbjerg1989}. Notably, a non-trivial behaivour is observed in concentrated regimes, where a secondary peak arises at $\phi \approx 50\%$. As shown in \citet{Park2010}, for a dilute suspension, the hydrodynamic diffusivity increases with a volume fraction before reaching a (primary) maximum and then significantly decreases in the semi-dilute regime up to $\phi = 20\%$. The current simulation results show that it continues to decrease and then reaches the local minimum at the volume fraction of about $35\%$. Interestingly, the hydrodynamic diffusivity starts to increase again in a concentrated regime until it reaches a local (secondary) maximum at a volume fraction of about $50\%$ and then drastically decreases as approaching random close packing. It is worth noting that the tracer diffusion (not shown) is found to show a qualitatively similar trend with figure \ref{fig2}$(a)$.

In order to further characterize the suspension dynamics, a statistical analysis is carried out for particle velocities. Since the velocity of the suspended particles only comes from the particle-particle interactions due to DIP, there is no net particle motion -- the mean particle velocity is exactly zero at every time step. Therefore, the particle velocity statistics can be readily characterized by the standard deviation of velocity distributions. Figure \ref{fig2}$(b)$ presents the standard deviation of particle velocity distributions in the transverse $(x)$ and field $(z)$ directions as a function of volume fraction. As seen in the figure, the magnitude of the velocity fluctuations increases with a volume fraction and reaches a maximum in the dilute regime ($\phi \approx 8\%$), and then decreases in semi-dilute suspensions. At $\phi \approx 30\%$, however, the velocity fluctuation starts to increase again, reaches a local peak at $\phi \approx 45\%$, and then decreases with a volume fraction. The velocity fluctuation indeed exhibits a similar trend to the diffusivity (figure \ref{fig2}$a$). In addition, it is seen that the fluctuation in the transverse direction starts to increase earlier (at $\phi \approx 25\%$) than that in the field direction (at $\phi \approx 30\%$). This observation suggests that suspension dynamics starts to change with regard to the preferred direction of particle motions at $\phi = 25\% \sim 30\%$.

The secondary peaks in the hydrodynamic diffusivity and velocity fluctuation are counter-intuitive since it is expected that the mobility of particles tends to decrease with concentration due to a consequence of strong excluded volume interactions. This distinct dynamics is most easily seen in the online movies for $\phi = 25\%$ and $45\%$, where the latter displays faster dynamics than the former. This non-trivial behaviour can result from several reasons. The clarification of the mechanisms contributed to the secondary peak will be illuminated as the paper proceeds.

\subsection{Suspension microstructure}
\begin{figure}
  \centerline{
  \includegraphics[width=4.8in]{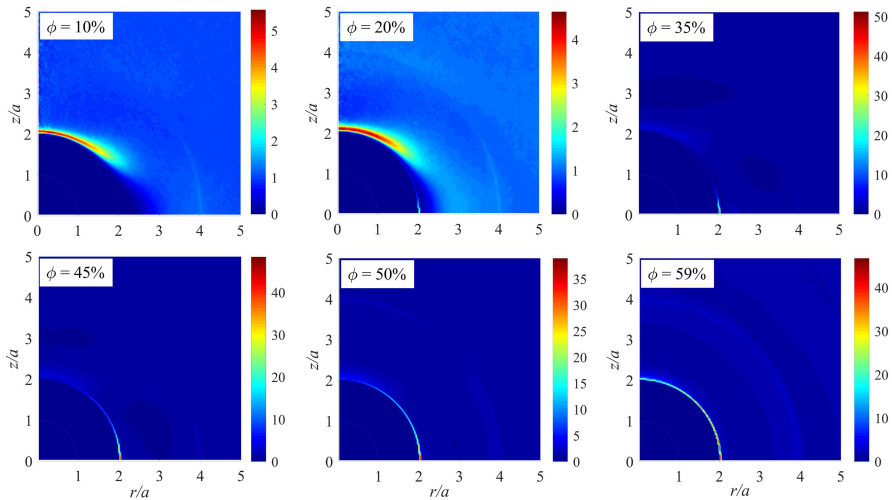}}
  \caption{Pair distribution functions in the suspensions undergoing dipolophoresis at six different volume fractions in cylindrical coordinates $(r, z)$, where $r^{2}=x^{2}+y^{2}$. A probe particle is located at the origin.}\label{fig5}
\end{figure}

The change in the suspension dynamics can be investigated quantitatively by calculating the pair distribution function $p(r, z)$, revealing the local microstructure of the suspension. This function provides the probability of finding the particle at a location $(r,z)$ in cylindrical coordinates ($r^{2}=x^{2}+y^{2}$) with respect to a probe particle located at the origin. The functions are shown in figure \ref{fig5} for various volume fractions. While dilute and semi-dilute suspensions ($\phi = 10\%$ and $20\%$) show the maximum probability of particles paired up in the field direction (near the pole of the particles), which has also been observed in suspensions of spherical and rod-like particles \citep{Park2010, Saintillan2006, Rose2009}, the peak location shifts from the pole to the equator in the concentrated regime ($\phi > 30\%$). The maximum probability region then seems to emanate and propagate toward the pole as further increasing the volume fraction. Finally, the peak region turns to entirely cover the particle surface at $\phi = 59\%$. The suspension microstructure undergoes a drastic change for $\phi > 30\%$, which might infer the non-trivial behaviours observed in the diffusivities and velocity fluctuations.

 \begin{figure}
  \centerline{
  \includegraphics[width=4.8in]{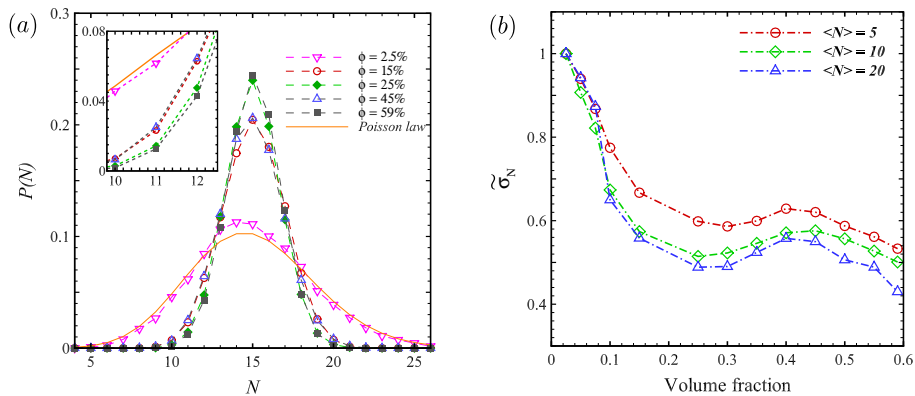}}
  \caption{$(a)$ Particle occupancy distributions for $\left \langle N \right \rangle=15$ at different volume fractions along with a Poisson distribution. $(b)$ The standard deviations of number-density fluctuations normalized with their maximum value as functions of volume fraction for three expected values $\left \langle N \right \rangle$ = 5, 10, and 20.}\label{fig6}
\end{figure}

In addition to the pair distribution function, particle occupancy statistics can be presented to characterize the number-density fluctuations for non-uniform microstructure on a larger length scale. To calculate the particle occupancy distribution, a cubic cell with a fixed volume $V$, which is calculated based on the fixed average number of the particle $\left \langle N \right \rangle = \phi V/V_p$, where $V_p$ is the volume of one sphere, is placed arbitrarily inside the simulation box. The cell will then contain an $N$ number of particles that can differ from the expected value $\left \langle N \right \rangle$. The probability or distribution of the number of particles $P(N)$ in the fixed cell volume presents the number-density fluctuation of the suspension. Figure \ref{fig6}$(a)$ shows the particle occupancy distribution for a cell containing $\left \langle N \right \rangle = 15$ particles on average. The curves include the occupancy distributions at steady-state for different volume fractions along with a Poisson distribution given by 
\begin{equation}
 P(N)=\frac{\left \langle N \right \rangle^{N} e^{-\left \langle N \right \rangle}}{N!}. \label{eq:poi}
\end{equation}   
It is shown that none of the distribution curves for $\phi > 15\%$ are captured by the Poisson distribution owing to strong excluded volume interactions in semi-dilute and concentrated suspensions as opposed to a dilute suspension at $\phi = 2.5\%$. To make a quantitative comparison, the standard deviation $\sigma_{N}$ of the particle occupancy distribution can be calculated for a range of volume fractions. As is the case in non-Poisson statistics \citep{Bergougnoux2009}, a power law of $\sigma_{N}$ as a function of $\left \langle N \right \rangle$ can effectively capture a number density deviation from the Poisson distribution $\sigma_{N} = \left \langle N \right \rangle^{1/2}$. It is found that all curves for $\phi > 15\%$ are well-fitted by a power law $\sigma_{N} \approx \left \langle N \right \rangle^{0.36}$, which is a quite substantial deviation from the Poisson law. It suggests a significant number density deviation from a random distribution at larger scales.

\begin{figure}
  \centerline{
  \includegraphics[width=4.8in]{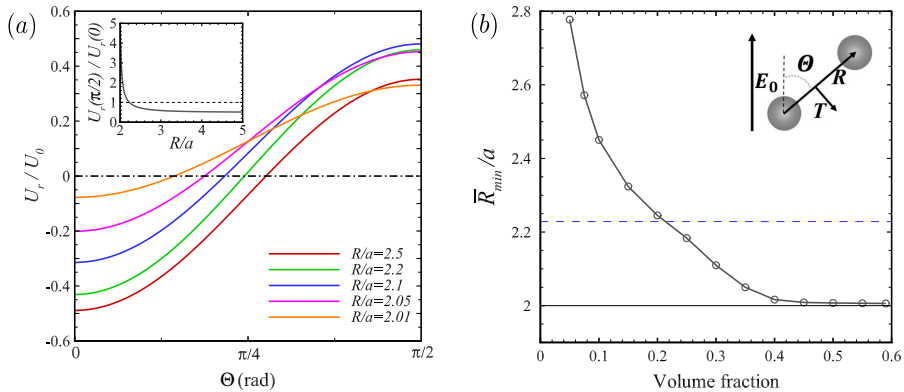}}
  \caption{$(a)$ Radial components of the total DIP relative velocity between two spheres as functions of the angle $\Theta$ between the direction of the electric field $\textbf{\textit{E}}_0$ and separation vector $\textbf{\textit{R}}$ for different separation distances. Velocities are scaled by $U_0 = (\epsilon a/\eta)E_0^2$. Inset: the magnitude ratio of the radial relative velocities in the transverse direction ($\Theta = \pi/2$) and in the field direction ($\Theta = 0$) as a function of separation distance $R/a$ -- the transverse velocity becomes dominant at $R/a \le 2.23$. $(b)$ The average minimum separation distance of particles in a suspension as a function of volume fraction along with a dashed line of $R/a = 2.23$. }\label{fig10}
\end{figure}

As seen in inset of figure \ref{fig6}($a$), the distributions at $\phi = 15\%$ and $45\%$ are broader compared to those at $\phi = 25\%$ and $59\%$. Specifically, it indicates that the number-density fluctuation initially decreases, then increases, and finally decreases with a volume fraction. To clearly illuminate this trend, figure \ref{fig6}$(b)$ shows the normalized standard deviation $\widetilde{\sigma}_N$ of number-density fluctuations as a function of volume fraction for three average numbers $\left \langle N \right \rangle$ = 5, 10, and 20. For all three cases, they share a similar trend with the diffusivities in figure \ref{fig2}$(b)$. After a local minimum in the semi-dilute regime ($\phi \approx 25\%)$, the standard deviation of the number-density fluctuation increases with a volume fraction in  concentrated suspensions and then decreases as approaching random close packing. This similar behaviour indicates that the fluctuation of the number of adjacent particles is strongly related to the hydrodynamic diffusion of the DIP suspension -- the larger number density fluctuations, the stronger hydrodynamic diffusion in the suspension.


\subsection{Dynamics under dipolophoresis in concentrated regimes}
We now attempt to illuminate the mechanisms behind the non-trivial behaviours in concentrated regimes. It has been known that in dense suspensions, particle interactions undergoing a lubricated-to-frictional transition provide a coherent mechanistic basis for the shear thickening in these suspensions \citep{Morris2018}. \citet{Mari2014} performed simulations for non-Brownian suspensions under shear and showed that a growing number of frictional contacts and its intermittent occurrence contribute to discontinuous shear thickening. Similarly, for concentrated DIP suspensions, particle contacts also appear to provide a coherent mechanistic basis for the non-trivial behaviours observed in concentrated regimes.

\begin{figure}
  \centerline{
  \includegraphics[width=4.8in]{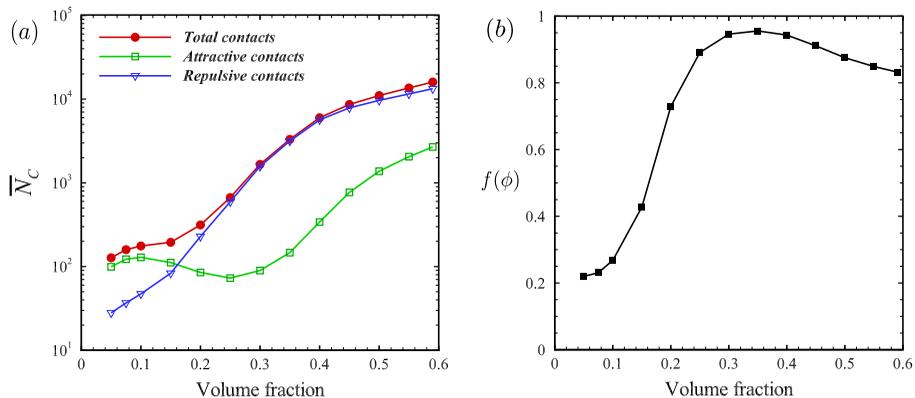}}
  \caption{$(a)$ The time-averaged number of total contacts, attractive contacts, and repulsive contacts as functions of volume fraction. $(b)$ The contact ratio $f$, which is the fraction of the number of repulsive contacts to total contacts, as a function of volume fraction.}\label{fig9}
\end{figure}

The suspension dynamics undergoing both DEP and ICEP or DIP is originated from pairing dynamics through electric and hydrodynamic interactions, contributing to particle contacts. For the case of a two-particle interaction, a radial component of the relative velocity $\textbf{\textit{U}} = \textbf{\textit{U}}_2 - \textbf{\textit{U}}_1$ can provide the nature of particle contacts. Figure \ref{fig10}$(a)$ shows the radial relative velocity $U_r = \textbf{\textit{U}}\cdot \hat{\textbf{\textit{R}}}$ for different separation distances between two particles as a function of the angle $\Theta$ between the axis of the two particles and the direction of the external field. The particles are attracted ($U_r < 0$) when aligned with the direction of the electric field, while they are repulsive ($U_r > 0$) when aligned in the transverse direction. At a large separation distance (e.g., $R/a = 2.5$), the field-direction velocity $U_r(\Theta = 0)$ is greater than the transverse-direction velocity $U_r(\Theta = \pi/2)$ and the large portion of the curve is on the attractive side. These results suggest that attractive contacts are likely to happen between two particles. As the particles come much closer (e.g., $R/a < 2.2$), however, the transverse-direction velocity is much greater the field-direction velocity and most of the curve is on the repulsive side. In such a case, the particles are strongly separated in the traverse direction and repulsive contacts are likely to happen if another particle is nearby -- this contact is highly expected to happen for concentrated suspensions. In the inset of figure \ref{fig10}($a$), the ratio of the transverse-direction velocity to the field-direction velocity is shown, where the transverse velocity becomes predominant at $R/a \le 2.23$. At $R/a = 2.01$, the repulsive velocity is almost four times the attractive velocity. To relate the changeover distance $R/a = 2.23$ to a characteristic separation distance of the suspension, the time-averaged minimum separation distance of all particles in a suspension is calculated at different volume fractions, as seen in figure \ref{fig10}$(b)$. At $\phi \approx 21\%$, the average minimum distance is about $R/a = 2.23$, at which the dominant direction of the pairing dynamics is expected to change from the field direction to the transverse direction. At higher volume fractions ($\phi > 45\%$), the average minimum distance of a suspension is almost $R/a = 2.01$, where repulsive forces are quite strong in the transverse directions. These forces are expected to strongly push particles each other along the transverse directions to make them contact nearby neighbours very often in concentrated suspensions.

Finally, we can make the link between particle contacts (excluded volume interactions) and the non-trivial behaviours in concentrated regimes. Given the contact direction, the particle contacts can be classified into attractive and repulsive classes with respect to an angle $\Theta$ at which a contact happens. We used the critical angle $\Theta_c = 0.11\pi$ at which $R/a = 2.075$ -- the issue of sensitivity to the chosen angle was tested using different angles at shorter distances, giving essentially identical results. The contact is called a repulsive contact if $\Theta > \Theta_c$ or an attractive contact if $\Theta < \Theta_c$. Figure \ref{fig9}$(a)$ shows the number of total, attractive, and repulsive contacts as a function of volume fraction. At a dilute regime ($\phi < 10\%$), there are more attractive contacts than repulsive contacts -- attractive interactions are the dominated mechanism for pairing dynamics. For $\phi > 16\%$, however, most of the contacts are from repulsive ones, indicating that repulsive interactions are the dominated mechanism of pairing dynamics that causes pushing particles to contact. To further manifest a dominance of repulsive contacts, we calculate the fraction of repulsive contacts $f(\phi)$, which is defined as the ratio of the number of repulsive contacts divided by the total number of contacts. Figure \ref{fig9}($b$) shows that there is a sharp increase in $f$ from $0.26$ ($\phi = 10\%$) to $0.90$ ($\phi = 25\%$). Then, almost 95\% of total contacts are repulsive in a range of $30\% < \phi < 40\%$. These strong, massive repulsive contacts are direct evidence as to the non-trivial behaviours in concentrated regimes -- particles tend to move very fast and meet others in the repulsive zone. The number of repulsive contacts is still around $85\%$ of the total contacts up to random close packing. These results are indeed related to the pair distribution functions, where the sharp peak probability region is located at the equator region and eventually covers the particle surface in concentrated regimes.

\section{Conclusion}\label{sec:conclusion}
We have presented large-scale numerical simulations of concentrated suspensions of ideally polarizable spheres undergoing dipolophoresis (the combination of dielectrophoresis and induced-charge electrophoresis). The previous model of \citet{Park2010} is adapted to simulate concentrated suspensions, to which the potential-free algorithm is added for excluded volume interactions to effectively prevent excessive particle overlaps. As reported in \citet{Park2010,park2011}, the induced-charge electrophoresis dominates the dynamics of such system and is responsible for the hydrodynamic diffusion and chaotic particle motion. In particular, the non-trivial behaviours are observed in concentrated regimes. In hydrodynamic diffusion, after $\phi \approx 35\%$ (local minimum), the hydrodynamic diffusivities starts to rise, reach a local peak at $\phi\approx 50\%$, and then decreases drastically with a volume fraction. A similar trend is observed for the velocity and number-density fluctuations, indicating that these fluctuations seem to enhance the hydrodynamic diffusion. Additionally, the suspension microstructure undergoes a significant change with a volume fraction, as measured by pair distribution functions and number-density fluctuations. The change in the local microstructure can tie into the non-trivial behaviours observed in suspension dynamics. 

As is the case in shear-thickening fluids, particle contacts due to DIP interactions are shown to provide the source of the non-trivial behaviours in concentrated regimes. Based on the pairing dynamics of two particles, the radial relative velocity $U_r$ in the transverse direction is found to become dominant over that in the field direction for $R/a \le 2.23$ and most of the $U_r$ curves are on the repulsive side. Consequently,  for a suspension of $\phi > 16\%$ where the averaged minimum separation distance becomes significantly small (i.e., $\bar{R}_{min}/a < 2.23$), most of the contacts are from repulsive contacts, indicating that the dominant mechanism of the particle contact is a strong repulsive interaction, causing particles to be  pushed to contact nearby ones in the transverse direction. The non-trivial behaviours are attributed to these strong, massive repulsive contacts, which are almost $95\%$ of total contacts in a range of $ 30\% < \phi < 40\%$ and approximately $85\%$ up to random close packing. These results are consistent with the change in pair distribution functions, where the peak shifts from the pole to the equator and eventually covers the particle surface. The follow-up work, aimed at linking the non-trivial behaviours and the rheology of such suspensions in concentrated regimes \citep{Tapia2017, Guazzelli2018, Butler2018}, is under investigation.

Lastly, this study appears to share similarities with active suspensions, especially pullers, thanks to the qualitatively same far-field fluid disturbances governed by a Stokes dipole (i.e., stresslet or symmetric force-dipole) \citep{Lauga2009,Saintillan2018}. In particular, the system studied is qualitatively identical to a suspension of spherical squirmers with a prescribed axisymmetric tangential velocity on its surface. However, the expected differences will be the magnitude and orientation of the surface (or slip) velocity, which may modify the relative importance of attractive and repulsive interactions between particles \citep{Ishikawa2006,Evans2011}. The non-trivial behaviour in concentrated regimes observed in this study may still arise in a squirmer or active suspension as a result of hydrodynamic interactions, depending on the magnitude and orientation of the surface velocity relative to ones of neighbouring particles. Such studies to investigate the generality of the non-trivial behaviour in concentrated regimes and a full comparison between inert and active particles \citep{Ishikawa2010} will be a subject of interesting future work.

\vspace*{0.1in}

The authors gratefully acknowledge the financial support from the Collaboration Initiative at the University of Nebraska and discussions with Jeffrey Morris and Michael Graham. The authors would also like to thank the anonymous referees for their valuable suggestions, especially similarities with a squirmer or active suspension. This work was completed utilizing the Holland Computing Center of the University of Nebraska, which receives support from the Nebraska Research Initiative.

\bibliographystyle{jfm}
\bibliography{DIP_concentrated_JFM}

\end{document}